\documentclass[aps,twocolumn,prl,showpacs,showkeys,floatfix]{revtex4}
\usepackage{epsfig}
\usepackage{isolatin1}
\usepackage{makeidx} \makeindex
\usepackage{amsmath} \usepackage{amssymb} \usepackage{amsthm}
\usepackage{mathrsfs} 

\begin{document}
\title {Dynamical suppression of telegraph and $1/f$ noise 
due to quantum bistable fluctuators}

\begin{abstract}
We study dynamical decoupling of a qubit from 
non gaussian quantum noise due to 
discrete sources, as 
bistable fluctuators and $1/f$ noise. We obtain analytic
and numerical results for generic operating point.
For very large pulse frequency, where dynamic decoupling 
compensates decoherence, we found universal behavior. 
At intermediate
frequencies 
noise can be compensated 
or enhanced, depending on the nature of the fluctuators 
and on the operating point.
Our technique can be applied to a larger class of non-gaussian environments.
\end{abstract}

\author{G. Falci} \affiliation{Dipartimento di Metodologie Fisiche e 
Chimiche (DMFCI), 
Universit\'a di Catania. Viale A. Doria 6, 95125 Catania (Italy) 
\& MATIS - Istituto Nazionale per la Fisica della Materia, Catania}
\author{A. D'Arrigo} \affiliation{Dipartimento di Metodologie Fisiche e 
Chimiche (DMFCI), Universit\'a di Catania. Viale A. Doria 6, 95125 Catania (Italy) 
\& MATIS - Istituto Nazionale per la Fisica della Materia, Catania}

\author{A. Mastellone} \affiliation{Dipartimento di Metodologie Fisiche e 
Chimiche (DMFCI), Universit\'a di Catania. Viale A. Doria 6, 95125 Catania (Italy) 
\& MATIS - Istituto Nazionale per la Fisica della Materia, Catania}
\author{E. Paladino} \affiliation{Dipartimento di Metodologie Fisiche e Chimiche (DMFCI), 
Universit\'a di Catania. Viale A. Doria 6, 95125 Catania (Italy) 
\& MATIS - Istituto Nazionale per la Fisica della Materia, Catania}

\email[email: ]{gfalci@dmfci.unict.it}


\maketitle
Controlling the dynamics of a complex quantum
system is at the heart of quantum 
information~\cite{kn:nielsen00}.
However 
in any real device
the computational variables 
entangle with the environment leading to
decoherence~\cite{kn:decoherence}. 
Bang-Bang (BB) control techniques
have been proposed as a way 
to achieve an effective decoupling from the 
environment~\cite{kn:viola98,kn:tombesi}. 
They may be operated by a sequence of strong
external pulses 
separated by a time $\Delta t$~\cite{kn:viola98}. 
For $\Delta t \to 0$ full 
decoupling~\cite{kn:viola98,kn:tombesi} of the unwanted
interactions is achieved.
The physics in this limit
is a manifestation of 
the quantum Zeno effect~\cite{kn:zeno,kn:viola98}.

In practice 
$\Delta t$ is finite
especially when full-power pulses are used. 
This imperfect decoupling is still well described 
by the Zeno limit if 
$\Delta t \ll \gamma^{-1}$, the typical time scale of 
the environment~\cite{kn:viola98,kn:tombesi}. 
If $\gamma$ is large one may argue that BB chops
noise and frequencies $\omega < 1/\Delta t < \gamma$ are
averaged out. This optimistic scenario could foresee  
applications to solid state coherent 
devices, where low-frequency noise~\cite{kn:oneoverf} is 
the major problem for quantum state 
processing~\cite{kn:paladino02,kn:falci00,kn:bc-exp}. 
Investigation of this point is one of the topics of this Letter,
where we study environments of dissipative 
quantum bistable fluctuators~\cite{kn:paladino02}. 

Recently decoupling from classical
Random Telegraph Noise (RTN) was studied 
in the Zeno limit, $\Delta t \ll \gamma^{-1}$~\cite{kn:wilhelm03}.
Gaussian noise with $1/f$ spectrum has also been 
studied~\cite{kn:lidar03} and 
decoupling for decreasing $\Delta t$ was found.
On the other hand in echo protocols, 
details of the structure and of the dynamics of a 
solid state discrete environment~\cite{kn:us-echo-class} 
may become important if the condition 
$\Delta t \ll \gamma^{-1}$ is not met.

We consider a qubit  
(${\cal H}_Q = -  \frac{\varepsilon}{ 2}  \, \sigma_z - 
\frac{\Delta}{2}  \, \sigma_x$) coupled to an impurity. 
The Hamiltonian is
\begin{equation}
\label{eq:hamiltoniana-impulsi-ideali}
{\cal H} \;=\; {\cal H}_Q \,-\, \frac{1}{2} \, \sigma_z \, 
\hat{E} \,+\, {\cal H}_E
\,+\, {\cal V}(t)
\end{equation}
The environment Hamiltonian
${\cal H}_E={\cal H}_d+ {\cal H}_T + {\cal H}_{B}$
describes an impurity level occupied by a localized 
electron, 
${\cal H}_d = \varepsilon_{c} d^\dagger d$, 
tunneling with amplitudes $T_{k}$ (${\cal H}_T =
\sum_{k} T_{k}  c_{k}^{\dagger} d + \mbox{h.c.}$)
to a fermionic band, described by
${\cal H}_{B} =\sum_k \varepsilon_{k} c^{\dagger}_{k}c_{k}$.
The charge in the impurity is coupled to the qubit, 
$\hat{E} = v \, d^{\dagger} d$. Control is operated as in Ref.~\cite{kn:viola98}, the 
external field ${\cal V}(t)$ 
being a sequence of $\pi$-pulses about $\hat{x}$.
This model may
describe charge noise due impurities 
close to a solid-state 
qubit~\cite{kn:paladino02,kn:zimmerman,kn:falci00,kn:theor}. 
The characteristic scale of the impurity is the 
switching rate 
$\displaystyle{\gamma=2 \pi {\cal N}(\epsilon_{c})|T|^2}$
(${\cal N}$ is the density of states of the fermionic band, 
$|T_{k}|^2 \approx |T|^2$).

This environment is non gaussian~\cite{kn:oneoverf}, a key 
feature to explain recent 
observations in Josephson qubits (splitting of spectroscopic peaks, 
beats in the coherent 
oscillations~\cite{kn:duty03}) due to
individual impurities close to the 
device. The observed $1/f$ noise 
is due to a set of such impurities~\cite{kn:bc-exp}.
We find that decoupling of this environment is sensitive to 
details of its dynamics. If pulses are not very frequent it
shows a rich variety of behaviors, 
suggesting 
that BB 
may also be used
for spectroscopy.

We operate with instantaneous pulses which do not 
modify the environment, 
the corresponding evolution operator being 
${\cal S}_P \approx 
i \sigma_x \otimes 1\!\!1_E$. 
The evolution operator 
of the Hamiltonian 
(\ref{eq:hamiltoniana-impulsi-ideali}) is
$[{\cal S}_P {\cal S}]^{2N}$, 
where ${\cal S} = 
\exp(- i {\cal H} \Delta t)$
with ${\cal V}(t)=0$ is 
the evolution between pulses.
Echo 
corresponds to $N=1$. The reduced density matrix (RDM) 
of the qubit is obtained
by tracing out the environment 
\begin{equation}
\rho(t) = 
\mathrm{Tr}_E 
\big\{[{\cal S}_P {\cal S}]^{2N} \, W(0) \, 
[{\cal S}^\dagger {\cal S}^\dagger_P]^{2N} \big\}
= {\cal E}_t[\rho(0)]
\label{eq:rdm}
\end{equation}
where $W(t)$ is the full density matrix 
and ${\cal E}_t[\cdot]$  is the quantum map~\cite{kn:nielsen00} 
associated to the reduced dynamics starting from a factorized
state, $W(0) = \rho(0) \otimes w_E$~\cite{kn:nota-1}. 

We may try to approximate Eq.(\ref{eq:rdm}) 
by a 
Bloch-Redfield master equation~\cite{kn:cohen}.
In this framework the  
environment remains in equilibrium and the map for the 
RDM
in the first
$\Delta t$
has the Lindblad form
${\cal E}_{\Delta t}[\rho(0)] \approx \exp({\cal L}\Delta t) \rho(0)$. 
The factorized structure of $W(t)$ is 
preserved if we apply 
pulses, so subsequent 
$\Delta t$ can be treated in the same way. After 
$t=2N \Delta t$
we get
\begin{equation}
{\rho}(t) \approx \big[{\cal P}\cdot 
\mathrm{e}^{{\cal L}\Delta t}\big]^{2N} 
{\rho}(0)
\label{eq:rdm-master}
\end{equation}
where ${\cal P}$ is the superoperator 
of the pulses. 
This approximation, which is correct 
for a weakly coupled 
and fast environment, 
yields that BB  does not affect the decay 
of the coherence. 
Of course BB decoupling
is effective only in situations where
memory effects are 
paramount, and the trace 
in Eq.(\ref{eq:rdm}) {\em must} be taken at the 
end of the protocol. In this cases we should go beyond 
the approximation Eq.(\ref{eq:rdm-master}). The possibility
we explore is to treat part of the environment 
on the same footing of the system~\cite{kn:paladino02}. 
We denote with 
$\rho(t)$ the RDM of the {\em qubit plus localized level}.
The system is now described by ${\cal H}_Q - {1 \over 2} \sigma_z \hat{E} + {\cal H}_d$
and we use the same steps leading
to Eq.(\ref{eq:rdm-master}).
The map 
$\rho(t+\Delta t)= \mathrm{e}^{{\cal L} \Delta t}\rho(t)$ 
is evaluated by a master equation~\cite{kn:paladino02},   
${\cal H}_T$ being the interaction and ${\cal H}_B$
the bath.
The RDM of the qubit $\rho^Q=\mathrm{Tr}_{d} [\rho(t)]$ 
is obtained by tracing out the 
localized level
{\em at the end} of the protocol.

We express $\rho(t)$ in the basis 
$| \theta_n^\pm \rangle = 
|\theta_n \pm \rangle | n\rangle$, where $| n\rangle$ ($n=0,1$) 
are eigenstates of $d^\dagger d$ 
and $|\theta_n \pm \rangle$ are the two eigenstates
of ${\cal H}_Q - (v/2)n \sigma_z$,
their energy splitting being
$\Omega_n = \sqrt{(\varepsilon+n v)^2 + \Delta^2}$.
We denote
$| a\rangle \equiv | \theta_0^+\rangle$,
$| b\rangle \equiv | \theta_0^-\rangle$,
$| c\rangle \equiv | \theta_1^+\rangle$ and
$| d\rangle \equiv | \theta_1^-\rangle$. 
In Refs.~\cite{kn:paladino02}
it was found that the impurity
remains in an unpolarized state,  
$\mathrm{Tr}_Q \{\rho(t)\} = 
\sum_{n=0,1} p_n(t) |n\rangle \langle n|$, 
if initially this was the case.
This simplifies the dynamics of $\rho_{ij}$: 
the only non vanishing entries are the four populations 
and the coherences $\rho_{ab}(t)$ and $\rho_{cd}(t)$ 
(with the conjugates). 
Thus we should diagonalize a $8 \times 8$ 
sub-matrix of $\cal L$. Using the 
representation of $\cal P$,
this is enough to find the approximate map 
Eq.(\ref{eq:rdm-master}) 
for a 
BB protocol, at all times.

If $\Delta = 0$,
the calculation can be carried out analytically. 
In absence of pulses 
$[{\cal H},\sigma_z]=0$, the populations of the qubit
do not relax while its coherences are given by 
$\langle \theta_0 +| \mathrm{Tr}_{d} [\rho(t)] | \theta_0 - \rangle
= 
\rho_{ab}(t) + \rho_{cd}(t)
$. This holds true also for an {\em even} number of pulses.
This symmetry 
further simplifies $\cal L$
leading
to independent evolutions of populations (subscript $p$) and  
coherences ($\phi$)
\begin{equation}
\label{eq:mappa-diss-singleBC-puro-deph}
\mathrm{e}^{{\cal L} t} \,\equiv\,
\left(\hskip-3pt
\begin{array}{cc}
\mathrm{e}^{{\cal L}_p  t} 
&
0
\\
0
&
\mathrm{e}^{{\cal L}_\phi  t} 
\end{array}
\hskip-3pt\right)
\quad ; \quad
\mathrm{e}^{{\cal L}_\phi t} \,\equiv\,
\left(\hskip-3pt
\begin{array}{cc}
	\mathrm{e}^{{\Gamma}_\phi t}
	&
	0
\\
	0
	&	\mathrm{e}^{{\Gamma}_\phi^* t}
	\end{array}
\hskip-3pt\right)
\end{equation}
where ${\cal L}_{p/\phi}$ are $4 \times 4$ matrices, whereas 
${\Gamma}_\phi$ is a $2 \times 2$ matrix acting on the vector 
${\rho}_\phi \equiv (\rho_{ab},\rho_{cd})$.
The pulse $\cal P$ is also diagonal in the $p-\phi$ indexes.  
In the $\phi$-subspace it is given by $I \otimes \sigma_x$, which allows to 
obtain the map for coherences $({\rho}_\phi,{\rho}_\phi\,\!^*)$ in an echo procedure
$$
[{\cal P} \mathrm{e}^{{\cal L}_\phi \Delta t} ]^2
\,\equiv\,
\left(\hskip-3pt
\begin{array}{cc}
	\mathrm{e}^{{\Gamma}_\phi^* \Delta  t} \mathrm{e}^{{\Gamma}_\phi \Delta t}
	&
	0
\\
	0
	& \mathrm{e}^{{\Gamma}_\phi \Delta t}	\mathrm{e}^{{\Gamma}_\phi^* \Delta t}
	\end{array}
\hskip-3pt\right)
$$
The 
``diagonal form'' implies that the game reduces to the two-component map 
${\rho}_\phi(t) = 
\big[\mathrm{e}^{{\Gamma}_\phi^* \Delta  t} \mathrm{e}^{{\Gamma}_\phi \Delta t}\big]^N 
{\rho}_\phi(0)$. This can be cast in a convenient form if the map 
$\mathrm{exp}({{\Gamma}_\phi  t})$ found in Refs.~\cite{kn:paladino02} is 
represented in $SU(2)$
\begin{eqnarray}
\label{eq:superop-coerenze-adiabatico-bang-su2}
{\rho}_\phi(t) = 
{\bigl[\mathrm{D} / |\alpha|^{2}\bigr]^N}\;
\mathrm{e}^{- \gamma \,N \Delta t \,+\, N\chi
\, {\sigma}_{\hat{\cal D}}
} \;
{\rho}_\phi(0)
\end{eqnarray}
Here $\alpha = [(1-i w)^2 - 2 i  
\,\delta p_{eq}\, g - g^2]^{1/2}$  determines 
the rates of the 
multi-exponential reduced  dynamics of the qubit,
the 
parameter 
$g=(\Omega_1-\Omega_0)/\gamma$
quantifies  non-gaussianity~\cite{kn:paladino02},
$\delta p_{eq}$ is
the equilibrium population difference of the fluctuator and
$w$ is related to the energy shifts produced by the band. 
Finally $\vec{\cal D}(\Delta t) \equiv ({\cal D}_x ,{\cal D}_y ,{\cal D}_z)$
and the quantities ${\cal D}_i(\Delta t)$ 
are easily found from the results   
in Refs.~\cite{kn:paladino02} 
(eg. ${\cal D}_0(\Delta t) =
|\alpha|^2  \bigl| \cosh \frac{\gamma \alpha \Delta t}{ 2}\bigr|^2 +
(1+ g^2 + w^2) \bigl|\sinh\frac{\gamma \alpha \Delta t}{ 2}\bigr|^2$)
and determine
$\mathrm{D}(\Delta t)=[{\cal D}_0 - |\vec{\cal D}|^2]^{1/2}$,  
$\chi(\Delta t) = \mathrm{arctanh} (|\vec{\cal D}|/{\cal D}_0)$ and
${\sigma}_{\hat{\cal D}} = \vec{\sigma} \cdot \vec{\cal D}/|{\cal D}|$.

Eq. (\ref{eq:superop-coerenze-adiabatico-bang-su2}) allows to discuss 
the dynamic decoupling of a quantum bistable fluctuator.
We expect a rich physics since this 
environment has distinctive
features 
depending
critically on $g$~\cite{kn:paladino02}.
Fast impurities ($g<1$) 
behave as an equivalent environment of harmonic oscillators 
in dephasing the qubit, whereas for  $g > 1$ a different 
physics emerges, 
dominated by memory effects, 
and decoherence depends strongly on details of the protocols.\begin{figure}[t]
\resizebox{64mm}{!}{
\includegraphics{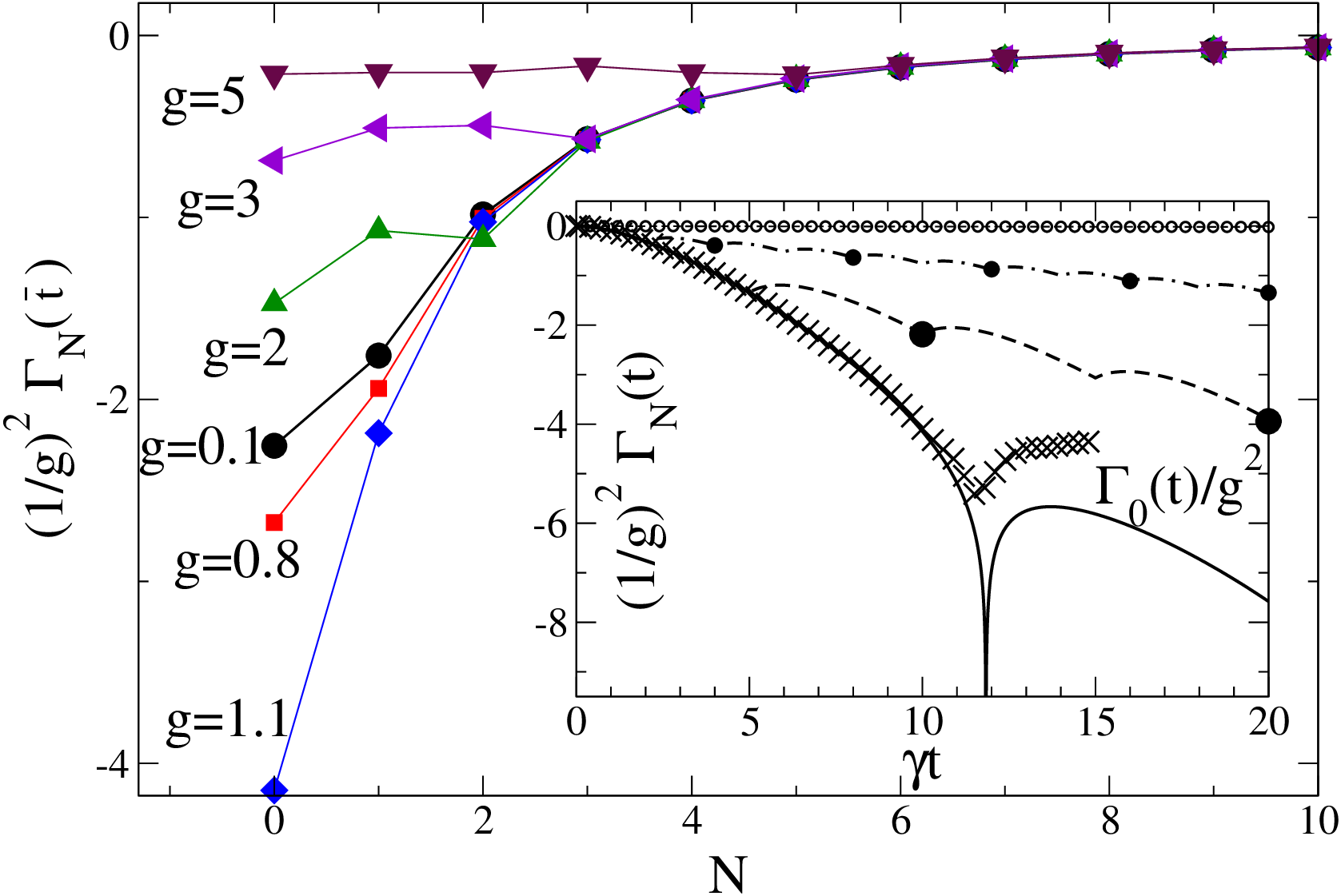}
}
\caption{For a fixed $\bar t=10 \gamma^{-1}$ we plot 
$\Gamma_N(\bar t)/g^2$ for 
BB procedures with $N$ echo pair of pulses. 
The parameter is $g \equiv v/\gamma$. 
$N=0$ corresponds to Free Induction Decay (FID)~\cite{kn:falci00,kn:bc-exp}. 
A gaussian environment
with the same power spectrum would give, for 
arbitrary $g$, the curve here labeled with $g=0.1$, 
since $\Gamma_N(t) \propto g^2$~\cite{kn:viola98}.
Inset: $\Gamma_N (t)$ for $g=1.1$ 
for different $\Delta t$ (lines with dots, $\gamma \Delta t=5, 2, 0.2$) 
are compared with the FID 
$\Gamma_0(t)$ (thick line)  and with results 
obtained by a stochastic Schr\"odinger 
equation (crosses) simulated with
a very efficient piecewise deterministic 
algorithm~\cite{kn:decoherence,kn:Falci-long04}. 
Simulations are not accurate at relatively long times
and in general they require a large statistics for the process
(we used $10^6$ realizations).
}  
\label{fig:pure-deph-rtn}
\end{figure}
We present (Fig.~\ref{fig:pure-deph-rtn}) the decay of 
the qubit coherences 
$
\Gamma_N(t) = \ln \Big| 
\frac{\rho_{ab}(t) + \rho_{cd}(t)}  
{\rho_{ab}(0) + \rho_{cd}(0)}
\Big|
$ 
in the limit of no back-action 
of the qubit on the impurity.  
This is obtained by letting $w = 0$~\cite{kn:paladino02}. 
At any fixed $\bar{t}=2N \Delta t$, 
$|\Gamma_N(\bar{t})|$ decreases monotonically when the pulse frequency 
$1/\Delta t$ increases, 
which shows that BB effectively suppresses RTN. For large frequencies, 
$1/\Delta t \gg \gamma$ (or $2N \gg \gamma t$),  $|\Gamma_N(\bar{t})|$ shows universal
behavior, scaling with 
$g^2$.
On the
other hand for $2N \ll \gamma t$ qualitative differences in the behavior are 
apparent for $g<1$ and $g>1$. Notice that 
for intermediate frequencies $1/\Delta t \lesssim \gamma$, the
regime of experimental interest, BB is 
still able to cancel part of the noise due to a fast fluctuator ($g<1$). 
For a slow fluctuator ($g>1$)  BB cancels the beats 
(minima in $\Gamma_0(t)$, inset of Fig.~\ref{fig:pure-deph-rtn})
in the coherent dynamics~\cite{kn:duty03} but  
besides this, it is weakly effective
against slow RTN, despite of 
the 
semiclassical arguments, 
because there is not much to cancel. 
Classical RTN 
causes also a systematic phase error 
which BB does not cancel~\cite{kn:Falci-long04}, 
but can be compensated otherwise.
Notice that the limit 
we discuss is the exact result for classical RTN,  
but Eq.(\ref{eq:superop-coerenze-adiabatico-bang-su2})
contains also the {\em quantum} dynamics
of the fluctuators, including  
the back-action of the qubit.
These results will be presented 
elsewhere~\cite{kn:Falci-long04}.

\begin{figure}[t]
\resizebox{61mm}{!}{
\includegraphics{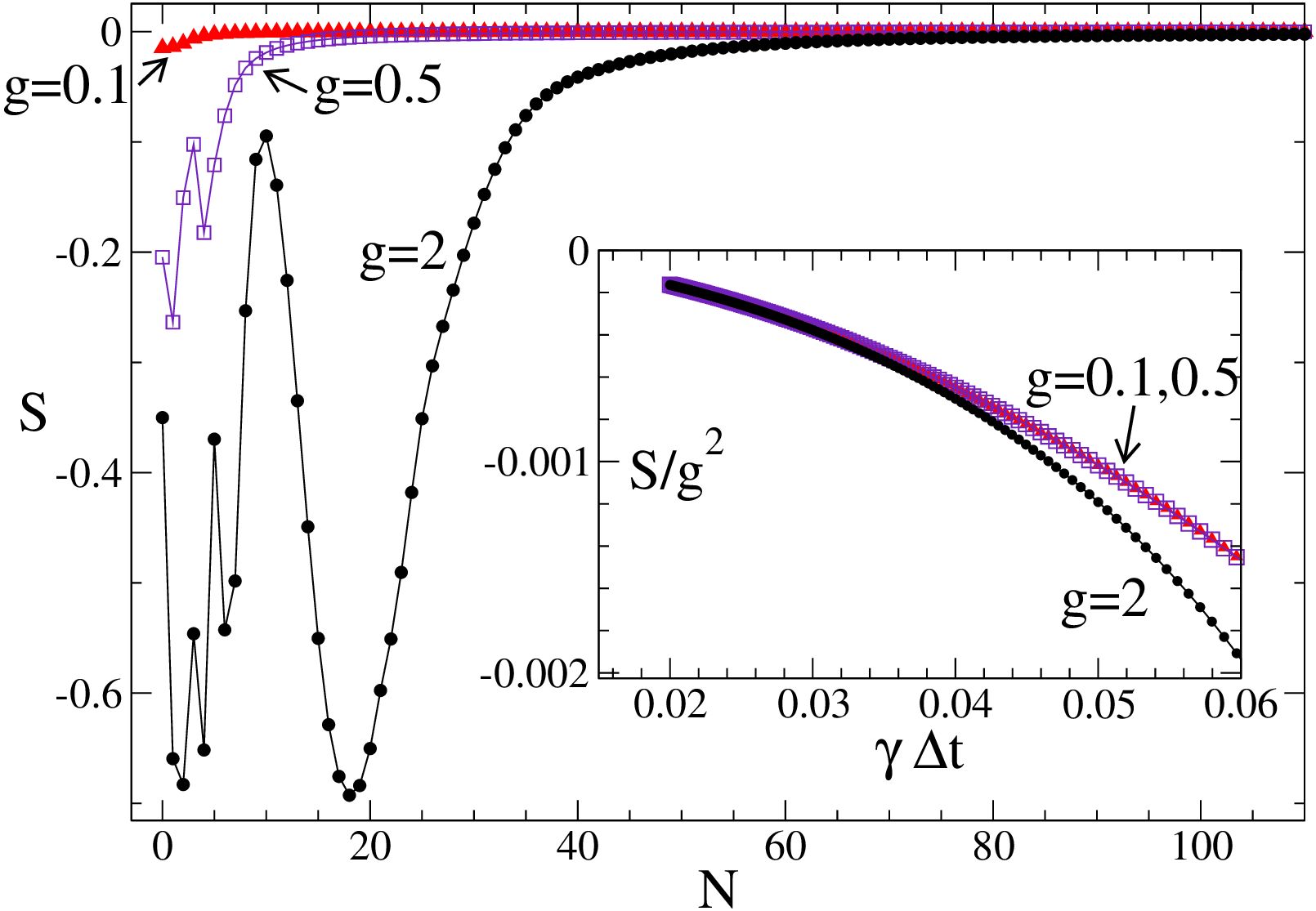}}
\caption{The purity 
$S= \ln[\mathrm{Tr}(\rho^Q)^2]$ at 
fixed $\bar t=8 \gamma^{-1}$
for protocols with $N$ echo pair of pulses. 
The parameter is 
$g = (\Omega_1 - \Omega_0)/\gamma$.
We take $\varepsilon = \Delta$, $v/\Omega = 0.2$ and start 
from an eigenstate of $\sigma_x$.
For $g=2$  (slow fluctuator), $S$ is non monotonic with 
$N$.
Fast fluctuators
($g=0.5$ and $g=0.1$) show a more regular behavior.
Inset: 
$S$ scales as $g^2$ for $N \gg 1$.
This regime of efficient decoupling is not 
easily met
for slow fluctuators ($g=2$ 
requires $\Delta t \lesssim \gamma^{-1}/25$).
}
\label{fig:gen-rtn}
\end{figure}

The physics for $\Delta \neq 0$ is even 
richer. We study the purity 
$S= \ln \mathrm{Tr}(\rho^Q)^2$, which gives 
deviations from unitary dynamics of the 
qubit~\cite{kn:nielsen00}.
Efficient decoupling, $S=0$, 
corresponds to localization 
in a ``Zeno subspace''~\cite{kn:zeno} 
which is a pure state. 
We study BB for {\em generic} $t$ and $\Delta t$
by 
diagonalizing  
$\exp({\cal L} t)$. Results (Fig.~\ref{fig:gen-rtn}) show that 
for frequent pulses decoupling is achieved, $S \approx 0$.
This agrees with the results of 
Ref.~\cite{kn:wilhelm03} for a $g<1$ impurity.
However decoupling slower impurities $g>1$,
requires comparatively large $N$. 
Universal behavior, 
$S \sim g^2$ is again found.
Instead for a smaller $N$ it may happen, 
especially for $g>1$  that $S$ is
not monotonic with $N$, including the possibility 
that 
the qubit decays {\em faster} 
than in absence of pulses~\cite{kn:viola98}. This is 
reminiscent of the {\em inverse} 
Zeno effect~\cite{kn:anti-zeno}, and it is
due to the 
complex coupled 
dynamics of qubit and impurity for 
$g>1$.

In order to treat $1/f$ noise 
we 
now extend our formalism to a ``multi-mode'' environment. 
We generalize 
the results at $\Delta = 0$ of 
Ref.~\cite{kn:viola98},  to an arbitrary (non-gaussian) 
environment. The Hamiltonian is of the general form 
Eq.~(\ref{eq:hamiltoniana-impulsi-ideali}). 
For the evolution between two pulses at $t_1$ and $t_2$
we use $[{\cal H},\sigma_z]=0$ and 
following the steps of Ref.~\cite{kn:viola98} we obtain 
the evolution operator at $t=2N \Delta t$ for a BB protocol, 
$
{\cal S}_{2N}(t) \,=\, 
\big[
\mathrm{e}^{- i ({\cal H}_E + {1 \over 2}  \sigma_z  \hat{E}) \Delta t}
\mathrm{e}^{- i ({\cal H}_E - {1 \over 2}  \sigma_z  
\hat{E}) \Delta t} 
\big]^N
$.
In the overall BB procedure $\sigma_z$ is conserved, so 
we need only off
diagonal entries of the RDM of the qubit,  
in the $\sigma_z$ basis
\begin{equation}
\label{eq:map-pure-dephasing}
\rho^Q_{\sigma \sigma^\prime}(t) =
\rho^Q_{\sigma \sigma^\prime}(0) \,
\mathrm{Tr}_E\Big\{ 
{\cal S}_{2N}(t|\sigma)  \, w_E \, {\cal S}_{2N}^\dagger(t|\sigma^\prime)
\Big\}
\end{equation}
where 
we assumed factorized initial conditions. Here 
${\cal S}_{2N}(t|\sigma) \,=\, 
\langle \sigma |{\cal S}_{2N}(t)|\sigma \rangle$
expresses
the conditional evolution of the environment under a well 
defined 
sequence of $\sigma=\pm 1$. 
The trace 
in Eq.~(\ref{eq:map-pure-dephasing}) 
factorizes if 
the environment is made of noninteracting ``modes'', if 
$\hat{E}$ are additive and if the initial $w_E$
is factorized.
If modes are 
oscillators one obtains the result of 
Ref.~\cite{kn:viola98}, which
 has been applied to a 
{\em gaussian} environment with $1/f$ 
spectral density~\cite{kn:lidar03}. 
This model may have 
limitations~\cite{kn:paladino02,kn:wilhelm03} in describing 
discrete noise sources of the solid-state, so we study
a more realistic model, the multi-mode version
of the Hamiltonian 
Eq.~(\ref{eq:hamiltoniana-impulsi-ideali}), 
${\cal H}_{E} \to \sum_\eta {\cal H}_{E\eta}$ and
$v \, d^\dagger d \to \sum_\eta v_\eta 
d_\eta^\dagger d_\eta$~\cite{kn:paladino02}. Each 
``mode'' is now a single impurity and 
we take a distribution $\propto \gamma^{-1}$ of the
individual switching rates
to produce $1/f$ noise~\cite{kn:oneoverf,kn:paladino02}. 

The contribution of each impurity to the 
suppression factor in 
Eq.(\ref{eq:map-pure-dephasing}) is calculated 
using the map Eq.(\ref{eq:rdm}).
The decay of the coherences is
$
\Gamma_N(t)
= \sum_\eta
\ln \Big|
{\rho^{(\eta)}_{ab}(t) + \rho^{(\eta)}_{cd}(t) \over 
\rho_{ab}(0) + \rho_{cd}(0)}
\Big|
$,  
where each $\rho^{(\eta)}_{ij}(t)$ is  
given by 
Eq.(\ref{eq:superop-coerenze-adiabatico-bang-su2}).
\begin{figure}[t]
\resizebox{68mm}{!}{
\includegraphics{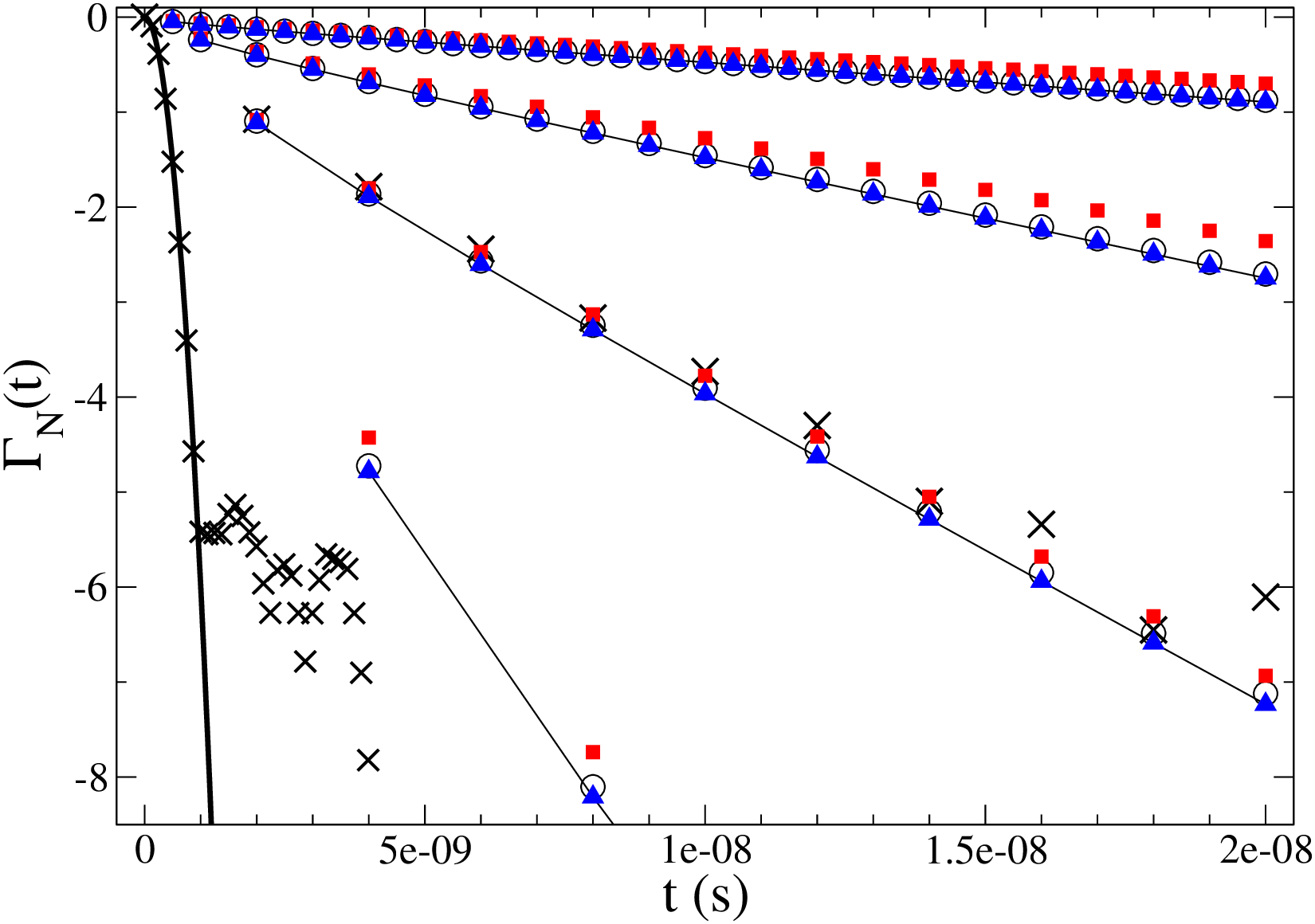}}
\caption{BB control of $1/f$ noise for 
$\Delta=0$. The analytic
$\Gamma_N(t)$ at $t=2N\Delta t$ (symbols - lines 
are guides for the eye) is compared with the evolution 
with no pulses (thick solid line). 
Noise is generated with $N_{fl}$
fluctuators with rates $\gamma_i$ distributed from 
$10^4 \,Hz$ to $10^{10} \, Hz$. Slower fluctuators are 
ineffective~\cite{kn:paladino02}. Noise level 
$\propto v^2 N_{fl}$ is {\em fixed} at a value typical of 
experiments in charge qubits: it is realized with 
coupling $v = 9.23 \cdot 10^q \,Hz$, for $q=6$
(full triangles), $7$ (circles) and $8$ (squares), with
$N_{fl}= 6 \cdot 10^{17-2q}$ scaled
accordingly.
Points for $q=6$ coincide
with results for gaussian noise with $1/f$ power spectrum.
Crosses are stochastic Schr\"odinger simulations with  
$10^5$ realizations of the $1/f$ process, for $q=7$.
}
\label{fig:pure-deph-oneoverf}
\end{figure}
Results in Fig.~\ref{fig:pure-deph-oneoverf} show that frequent pulses (curves with
many symbols) suppress decoherence. 
Under pulsed control $\Gamma_N(t)$ changes from
$\propto t^2$ to $\propto t$, i.e. it is
described by a rate depending on $\Delta t$,
as in the Zeno effect. 
For noise levels typical of experiments with charge qubits 
(Fig.~\ref{fig:pure-deph-oneoverf}) the pulse rate for 
substantial recovery is practically independent 
on $v$. Thus the criterion for efficient 
decoupling proposed in Ref.~\cite{kn:prefiche} is not 
effective in this regime. The situation may change if a broad distribution
of couplings is considered~\cite{kn:us-echo-class}.
The physics is richer for 
$\Delta \neq 0$~\cite{kn:Falci-long04,kn:prefiche}
and compensation of 
$1/f$ noise is non monotonic for decreasing $\Delta t$, 
as for a single impurity.

BB suppression of 
noise (RTN and $1/f$) due to quantum fluctuators
is an example of
general situations where 
a ``structured'' environment is involved. Indeed 
the qubit interacts mainly with the impurity, which 
is a ``quantum filter'' modulating the 
noise from the band. We treat 
this filter on the same footing of the qubit.

Universal behavior in terms of the 
scaling parameter $g$ 
for very frequent pulses ($\Delta t \ll \gamma^{-1}$) 
indicates 
that when decoupling is effective details of the 
environment are unimportant. 
Instead in the experimentally relevant case of finite $N$ 
($\Delta t \gtrsim \gamma^{-1}$),
the different physics of slow ($g >1$) and fast ($g<1$) 
fluctuators manifests itself, and may give 
rise to decoupling and/or 
to enhancement of decoherence.
This picture, unexpected in the naive  
description of BB, is reminiscent of
the inverse Zeno effect~\cite{kn:anti-zeno}. 
The BB scheme we discussed prevents 
decoherence but freezes part of the dynamics. More
complicated schemes may 
also allow to perform computation~\cite{kn:control}. 
The rich physics we find 
suggests that BB may be used to extract informations on the 
environment, e.g. for 
$1/f$ noise at otherwise inaccessible frequencies.
Results discussed here are exact 
for classical RTN and $1/f$ noise, but the formulas we
presented have a broader validity:
we also studied~\cite{kn:Falci-long04}
the back-action of the qubit on the fluctuator and 
$1/f$ noise at general bias point, confirming the
qualitative picture of this work.
We finally stress that our results 
apply to other sources of 
discrete noise, as flux 
or critical current noise in flux qubits.

We thank 
P. Facchi, R. Fazio, F. Petruccione  and 
F. Wilhelm for several comments and discussions.


\begin{thebibliography}{99}
\bibitem{kn:nota-1} Entangled initial states obtained 
by preparing 
a factorized state at an earlier time $t_0 <0$ can be 
treated on the same footing.
\end{thebibliography}
\end{document}